\documentclass[aps,prb,twocolumn,superscriptaddress,showpacs]{revtex4-1}
\pdfoutput=1

\usepackage{epstopdf}
\usepackage{epsfig}
\usepackage{amsmath}
\usepackage{amssymb}
\usepackage[colorlinks=true,citecolor=blue,linkcolor=blue]{hyperref}

\begin{document}

\title{Doping the Kane-Mele-Hubbard model: A Slave-Boson Approach}

\author{Jun Wen}
\email{jwen@physics.utexas.edu}
\affiliation{Department of Physics, The University of Texas at Austin, Austin, TX 78712, USA}

\author{Mehdi Kargarian}
\email{kargarian@physics.utexas.edu}
\affiliation{Department of Physics, The University of Texas at Austin, Austin, TX 78712, USA}

\author{Abolhassan Vaezi}
\affiliation{School of Physics, Institute for Research in Fundamental Sciences, IPM, Tehran, 19395-5531, Iran}
\affiliation{Department of Physics, Sharif University of Technology, Tehran 11155-9161, Iran}

\author{Gregory A. Fiete}
\affiliation{Department of Physics, The University of Texas at Austin, Austin, TX 78712, USA}

\begin{abstract}
We study the Kane-Mele-Hubbard model both at half-filling and away from half-filling using a slave-boson mean-field approach at zero temperature. We obtain a phase diagram at half-filling and discuss its connection to recent results from quantum Monte Carlo, cellular dynamical mean field, slave-rotor, and $Z_2$  mean-field studies. In particular, we find a small window in parameter space where a spin liquid phase with gapped spin and charge excitations reside. Upon doping, we show the spin liquid state becomes a superconducting state by explicitly calculating the singlet pairing order parameters. Interestingly, we find an ``optimal" doping for such superconductivity. Our work reveals some of the phenomenology associated with doping an interacting system with strong spin-orbit coupling and intermediate strength electron-electron interactions.

\end{abstract}
\date{\today}

%\pacs{71.10.Fd,71.10.Pm,73.20.-r}

%71.10.Fd 	Lattice fermion models (Hubbard model, etc.)
%72.80.Ga Transition-metal compounds, electrical conductivity of
% 03.65.Vf Topological phases (quantum mechanics)
%71.10.Pm Fermions in reduced dimensions (anyons, composite fermions, Luttinger liquid, etc.)
%73.20.-r 	Electron states at surfaces and interfaces

\maketitle

\section{Introduction}
Recent years have seen growing interest in topological band insulators (TBI).\cite{Kane:rmp10,Moore:nat10,Zhang:pt10} While in the fractional quantum Hall effect the Coulomb interaction is necessary for the topological order, the concept of TBIs can be readily appreciated in the framework of noninteracting Bloch band theory where spin-orbit coupling is responsible for a possibly nontrivial $Z_2$ topological order.\cite{Kane:rmp10,Moore:nat10,Zhang:pt10} A TBI has a gap in the bulk excitation spectrum and time-reversal symmetry protected gapless modes on the boundary. Because in nature all materials possess Coulomb interactions, understanding the role of  interactions is one of the fundamental challenges in the field of topological insulators.

One can ask if it is possible for interactions to induce TBIs. The answer is affirmative. Indeed, there have been a number of works that address this question with different models that contain {\em no intrinsic spin-orbit coupling}. For example, Raghu \emph{et al.}\cite{Raghu:prl08}  showed it is possible to have an interaction-driven TBI with spontaneously broken SU(2) symmetry (with spontaneously generated spin-orbit coupling) from an extended Hubbard model on the honeycomb lattice.  This idea has been successfully applied to the kagome lattice\cite{Wen:prb10}  and the decorated honeycomb lattice\cite{Wen:prb10} in 2D  among others,\cite{Sun:prl09,Yi:prb09,Liu:prb10,FZhang:prl11,Ruegg11_2,Wang11,Yang11} and the diamond lattice in 3D.\cite{Yi:prb09} The key is to have the correct amount of ``generalized" spin-orbit coupling that originates from the Hartree-Fock mean-field decoupling of the interaction terms on nearby sites.

Another equally important question is the fate of TBIs with intrinsic spin-orbit coupling upon the inclusion of Coulomb interaction. On one hand, by the argument of adiabatic continuity, it is argued that a TBI should be stable to weak interactions as long as the bulk gap is not closed.\cite{Kane:2005a,Kane:2005b} However, when interactions grow too strong, one has a good reason to believe that spin-charge separation develops and Mott physics will appear.\cite{Pesin:np10} In this regime, one expects that a slave particle approach which starts with an explicit decomposition of the electron into charge and spin degrees of freedom would qualitatively capture the physics of the interactions. Indeed, back in 2008 Young \emph{et al.}\cite{Young:prb08} employed a slave-rotor mean-field approach to study a double layer honeycomb lattice where a fractionalized quantum spin Hall (FQSH) effect could be found. A FQSH state differs from a quantum spin Hall state in that neutral spinons instead of physical electrons carry a nontrivial $Z_2$ topology. As a result, a gapless spinon excitation is guaranteed to appear along the edge. Applying similar methods, others\cite{Rachel:prb10,Kou:prb11} used the same approach to study the Kane-Mele-Hubbard model [our Eq.\eqref{KMH-model}] on the single-layer honeycomb lattice and concluded that this phase could be stabilized if the two dimensional U(1) gauge field is screened by an additional metallic layer so that the gauge fluctuations are suppressed.

In three dimensional systems, Pesin and Balents\cite{Pesin:np10} studied heavy transition-metal oxides on the pyrochlore lattice and proposed a three dimensional counterpart of the FQSH, termed as a ``topological Mott insulator" (TMI).  A TMI is one example of a U(1) spin liquid (SL) in three dimension and is believed to be more stable to gauge fluctuations than its two dimensional counterpart.\cite{Witczak-Krempa:prb10} Later, Kargarian \emph{et al.}\cite{Kargarian:prb11} extended Pesin and Balent's results and investigated the interplay between interactions and distortion in the same system. Based on these works, it may appear that the concept of the FQSH in two dimensions and the TMI in three dimensions depends crucially on the slave-rotor approach, which by its construction transfers the topology of physical electrons to neutral spinons and makes access to fractionalized states possible.
%In the classification of interacting TBI, Wang \emph{et al.}\cite{Wang:prl10} has generalized the topological order parameter for interacting topological insulators and expressed it in terms of the full Green's functions of the interacting system.

The extent to which a slave-rotor mean-field approach is reliable can be checked with more controlled numerical simulations. Recent quantum Monte Carlo and cellular dynamical mean field studies have shed light on the weak and intermediate interaction regimes in two dimensions.\cite{Meng:nat10,Assaad:prl11,Wu:10,Imada:prb11,Hur:11}
In a pioneering quantum Monte Carlo study, Meng \emph{et al.}\cite{Meng:nat10} investigated the Hubbard model on the honeycomb lattice at half-filling and discovered the existence of a gapped spin liquid in a small window in the intermediate interaction regime ($3.5t< U<4.3t$). Later, spin-orbit coupling was included and the spin liquid phase was found to be stable for small spin-orbit coupling \cite{Assaad:prl11} and for finite temperatures.\cite{Hur:11} At half-filling, the above quantum Monte Carlo studies are free of the sign problem and considered to be accurate. Of particular interest is the nature of the spin liquid, which has been addressed in a number of works.\cite{Wang:prb10,Sondhi:10,Vaezi:10,Lu_10_1,Lu_10_2}  Very recent work has indicated that beyond a critical interaction strength and spin-orbit coupling strength (larger than that explored in quantum Monte Carlo) yet another novel phase may appear with fractionized excitations and a non-trivial ground-state degeneracy\cite{Ruegg11} and attention has been drawn to transition metal oxide interfaces.\cite{Ruegg11_2,Wang11,Yang11,Xiao11}

In this paper we aim to better understand the intermediate interaction regime where a gapped SL phase appears. We are particularly interested in the fate of the SL\cite{Fiete:11} upon doping. This is a regime where quantum Monte Carlo simulations suffer from the sign problem\cite{Troyer:prl05} and the slave-rotor mean-field approach may encounter severe limitations\cite{Georges:prb02} leaving few tools available for its study.  We will follow Ref.~[\onlinecite{Vaezi:10}] and use a generalized U(1) slave-boson mean-field approach to study the cases of half-filling {\em and} doping. Such an approach has been widely used in doped $t$-$J$ models in the context of high temperature superconductivity.\cite{Wen:rmp06}  {\em We stress that we do not expect the slave-boson mean-field approach to represent a good solution to the Kane-Mele-Hubbard model in all regimes.} Instead, we argue that it gives a reasonably good description of the gapped SL at intermediate regime (based on a quantitative comparison with QMC and CMDFT) and its transition to a superconducting state upon doping. For a general review of Hubbard model, we refer interested readers to Ref.~[\onlinecite{HubbardBook:1995}].

This paper is organized as follows.  In Sec. \ref{sec:slave} we introduce the slave-boson representation for the Kane-Mele-Hubbard model.  In Sec. \ref{sec:results} we describe our slave-boson mean-field results for the cases of half-filling and doping.  Finally, in Sec.
\ref{sec:conclusions} we give the main conclusions of this work. In App. \ref{app:formulas} we provide some lengthy self-consistency formulas used to obtain our results.

\section{The slave-boson approach}
\label{sec:slave}
We start with the Kane-Mele-Hubbard model on the honeycomb lattice,
\begin{equation}
\label{KMH-model}
H=-t \sum_{\langle ij \rangle} c^{\dag}_{i,\sigma}c_{j,\sigma}+U\sum_{i} n_{i\uparrow}n_{i\downarrow} + i\lambda_{SO}\sum_{\langle\langle i,j \rangle\rangle,\sigma}\sigma\nu_{ij}c^{\dag}_{i,\sigma}c_{j,\sigma}
\end{equation}
where $t$, $U$, and $\lambda_{\rm{SO}}$ are the nearest neighbor hopping energy, the strength of the on-site repulsion, and  the second-neighbor spin-orbit coupling strength, respectively. Here $c_{i\sigma}$ ($c_{i\sigma}^{\dag}$) annihilates (creates) an electron with spin $\sigma$ on site $i$ and $\nu_{ij}=\pm 1$ depending on if the electron makes as ``right" or ``left" turn when going from $i$ to $j$.\cite{Kane:prl05,Kane_2:prl05}

The general U(1) slave-boson approach decomposes an electron operator into a bosonic operator that carries the charge degree of freedom and a fermionic spinon operator that carries the spin degree of freedom:\cite{Barnes:jpf76,Coleman:prb84,Zou:prb88,Wen:rmp06,Vaezi:10}
\begin{equation}
\label{slave-boson-decomposition}
c^{\dag}_{i,\sigma}=f^{\dag}_{i,\sigma}h_i+\sigma d^{\dag}_i f_{i,-\sigma},
\end{equation}
where $i$ is the site index, and $h_i$ and $d_i$ are the bosonic holon operator and the bosonic doublon operator, respectively. Such a decomposition makes the idea of spin-charge separation explicit and one expects that it will describe the physics of intermediate (and possibly strong) interactions reasonably well.

There are four states, $|0\rangle$,$|\uparrow\rangle$,$|\downarrow\rangle$, and $|\uparrow \downarrow\rangle$, at each site.  Each state can be thought to have some new particle operator acting on some vacuum state: $|0\rangle = h^{\dag} |vac\rangle $,  $|\uparrow \rangle = f^{\dag}_{\uparrow} |vac\rangle $, $ |\downarrow\rangle = f^{\dag}_{\downarrow} |vac\rangle $ and $ |\uparrow \downarrow\rangle = d^{\dag}|vac\rangle $.  Physically, one can think of $h^{\dag}_i h_i$ as the number of empty occupancies at site $i$, $f^{\dag}_{i\sigma}f_{i\sigma}$ the single occupancy with spin $\sigma$, and $d^{\dag}_i d_i$ the double occupancy. One can show that Eq.~(\ref{slave-boson-decomposition}) guarantees that the matrix elements of physical states are correct. The completeness of the basis implies the constraint
\begin{equation}
\label{constrain-completeness}
h^{\dag}_i h_i+\sum_\sigma f^{\dag}_{i,\sigma}f_{i,\sigma}+d^{\dag}_i d_i=1,
\end{equation}
which also preserves the anticommunication relations of $c_{i,\sigma}$ and $c^{\dag}_{i,\sigma}$. This constrain can be enforced with a Lagrange multiplier $\lambda_i$ in the Hamiltonian.

There is also another constraint related to the filling fraction of electrons: $
c^{\dag}_{i\sigma}c_{i\sigma}=f^{\dag}_{i\sigma}f_{i\sigma}+d^{\dag}_id_i $
where some extra terms which have zero matrix elements in the physical states have been thrown away.\cite{Barnes:jpf76,Coleman:prb84,Zou:prb88,Wen:rmp06,Vaezi:10} Therefore,
\begin{equation}
\label{eq:doping}
\sum_{\sigma} \langle c^{\dag}_{i\sigma}c_{i\sigma} \rangle=1+\langle d^{\dag}_id_i \rangle - \langle h^{\dag}_ih_i\rangle\equiv 1+x,
\end{equation}
where $x$ is the electron doping. We can incorporate the constraint \eqref{eq:doping} by another Lagrange multiplier $\mu_i$ in the Hamiltonian.

With the slave-boson representation described above, the Kane-Mele-Hubbard model can be written as
%\begin{widetext}
%\begin{eqnarray}
%\label{KMH-model3}
%H=U\sum_i d^{\dag}_id_i-t \sum_{\langle ij \rangle} %\left[\chi^f_{ij}\chi^b_{ji}+\Delta^{f\dag}_{ij}\Delta^{b}_{ij}+h.c.\right ] %\nonumber \\
%+ \lambda_{\rm{SO}} \sum_{\langle\langle ij \rangle\rangle} \left[\chi^{f\prime}_{ij}\chi^{b\prime}_{ji}+\Delta^{f\dag\prime}_{ij}\Delta^{b\prime}_{ij}+h.c.\right ] \nonumber \\
% - \sum_{i}\lambda_i\left (h^{\dag}_i h_i+\sum_\sigma f^{\dag}_{i\sigma}f_{i\sigma}+d^{\dag}_i d_i-1 \right) \nonumber \\
%-\sum_{i}\mu_i\left(  d^{\dag}_id_i - h^{\dag}_ih_i -x \right ),
%\end{eqnarray}
%\end{widetext}

\begin{widetext}
\begin{eqnarray}
\label{KMH-model3}
H=-t \sum_{\langle ij \rangle} \left[\chi^f_{ij}\chi^b_{ji}+\Delta^{f\dag}_{ij}\Delta^{b}_{ij}+h.c.\right ]+ \lambda_{\rm{SO}} \sum_{\langle\langle ij \rangle\rangle} \left[\chi^{f\prime}_{ij}\chi^{b\prime}_{ji}+\Delta^{f\dag\prime}_{ij}\Delta^{b\prime}_{ij}+h.c.\right ]+U\sum_i d^{\dag}_id_i \nonumber \\
 - \sum_{i}\lambda_i\left (h^{\dag}_i h_i+\sum_\sigma f^{\dag}_{i\sigma}f_{i\sigma}+d^{\dag}_i d_i-1 \right)
-\sum_{i}\mu_i\left(  d^{\dag}_id_i - h^{\dag}_ih_i -x \right ),
\end{eqnarray}
\end{widetext}
where the following order parameters are defined for nearest neighbor (NN) sites $\langle ij \rangle$: $\chi^{f}_{ij}=\sum_\sigma f^{\dag}_{i\sigma}f_{j\sigma}$, $
\chi^{b}_{ij}=h^{\dag}_{i}h_{j}-d^{\dag}_{i}d_{j}$,
$\Delta^{f}_{ij}=\sum_\sigma \sigma f_{i-\sigma}f_{j\sigma}$, $\Delta^{b}_{ij}= d_{i}h_{j}+h_id_j$ and for the next nearest neighbor (NNN) sites $\langle\langle ij \rangle\rangle$: $\chi^{f\prime}_{ij}=\sum_\sigma i\nu_{ij} \sigma f^{\dag}_{i\sigma}f_{j\sigma} $, $\chi^{b\prime}_{ij}=h^{\dag}_{i}h_{j}-d^{\dag}_{i}d_{j}$,
$\Delta^{f\prime}_{ij}=\sum_\sigma i\nu_{ij} f_{i-\sigma}f_{j\sigma} $ and
$\Delta^{b\prime}_{ij}= d_{i}h_{j}+h_id_j $.
%\begin{widetext}
%\begin{eqnarray}
%\label{KMH-model2}
%H=U\sum_i d^{\dag}_id_i-t \sum_{\langle ij \rangle}[f^{\dag}_{i\sigma}f_{j\sigma}h_ih^{\dag}_j+f_{i-\sigma}f^{\dag}_{j-\sigma}d^{\dag}_id_j
%+\sigma f_{i-\sigma}f_{j\sigma}d^{\dag}_ih^{\dag}_j+\sigma f^{\dag}_{i\sigma}f_{j-\sigma}^{\dag}h_id_j] - \sum_{i}\lambda_i(h^{\dag}_i h_i+\sum_\sigma f^{\dag}_{i\sigma}f_{i\sigma}+d^{\dag}_i d_i-1)\nonumber \\ +  i\lambda_{\rm{SO}} \sum_{\sigma,\langle\langle ij \rangle\rangle}\sigma\nu_{ij}[f^{\dag}_{i\sigma}f_{j\sigma}h_ih^{\dag}_j+f_{i-\sigma}f^{\dag}_{j-\sigma}d^{\dag}_id_j
%+\sigma f_{i-\sigma}f_{j\sigma}d^{\dag}_ih^{\dag}_j+\sigma f^{\dag}_{i\sigma}f_{j-\sigma}^{\dag}h_id_j]
%-\sum_{i}\mu_i(  h^{\dag}_ih_i -d^{\dag}_id_i-x)
%\end{eqnarray}
%\end{widetext}

We proceed with a mean-field approximation in which the spinon part and the boson part decouple from each other. We will restrict ourselves to a search for phases preserving translational symmetry. The simplest phase is the one that does not break any symmetry of the Hamiltonian.  In this case each type of order parameter  does not depend on the site indices.   For example,
 $\langle\chi^{b}_{ij}\rangle= \chi_{b}$
%\begin{eqnarray}
%\langle\chi^{b}_{ij}\rangle= \chi_{b} \\
%\langle\chi^{f}_{ij}\rangle= \chi_{f} \\
%\langle\Delta^{f}_{ij}\rangle=\Delta_{f} \\
%\langle\Delta^{b}_{ij}\rangle=\Delta_{b}
%\end{eqnarray}
for any $\langle ij \rangle$.
We then have $ H_{MF}=H_f+H_b+H_{const} $
where
\begin{widetext}
\begin{align}
\label{Hf}
H_f &= \sum_{\textbf{k}\sigma} \left[ -t\chi_{b}g(\textbf{k})f^{\dag}_{\textbf{k}A\sigma}f_{\textbf{k}B\sigma}
 -t\Delta_b g(\textbf{k})\sigma f^{\dag}_{\textbf{k}A\sigma}f^{\dag}_{-\textbf{k}B-\sigma}
 + \lambda_{\rm{SO}}\chi_b^{\prime} \sigma (f^{\dag}_{\textbf{k}A\sigma}f_{\textbf{k}A\sigma}
 -f^{\dag}_{\textbf{k}B\sigma}f_{\textbf{k}B\sigma})g_1(\textbf{k}) \right ]\nonumber \\
 &-\lambda \sum_{\textbf{k}\sigma\alpha}f^{\dag}_{\textbf{k}\alpha\sigma}f_{\textbf{k}\alpha\sigma} +\lambda_{\rm{SO}}\Delta_b^{\prime}\sum_{\textbf{k}}\left(f^{\dag}_{\textbf{k}A\uparrow}f^{\dag}_{-\textbf{k}A\downarrow}
 -f^{\dag}_{\textbf{k}B\uparrow}f^{\dag}_{-\textbf{k}B\downarrow} \right) + h.c.,
\end{align}
\begin{align}
\label{Hb}
H_b &= \sum_{\textbf{k}\alpha} \left [ \left( U-\lambda-\mu-\lambda_{\rm{SO}}\chi_f^{\prime}g_2(\textbf{k}) \right ) d^{\dag}_{\textbf{k}\alpha}d_{\textbf{k}\alpha}+\left(\mu- \lambda+ \lambda_{\rm{SO}}\chi_f^{\prime}g_2(\textbf{k})\right ) h^{\dag}_{\textbf{k}\alpha}h_{\textbf{k}\alpha} +
\lambda_{\rm{SO}}\Delta_f^{\prime}g_2(\textbf{k})d_{\textbf{k},\alpha}h_{-\textbf{k},\alpha}
\right ]\nonumber \\
&-\sum_{\textbf{k}} \left [ t\chi_{f}(h^{\dag}_{kA}h_{kB}
-d^{\dag}_{\textbf{k}A}d_{\textbf{k}B})g(\textbf{k})+t\Delta_{f}(d_{\textbf{k}A}h_{-\textbf{k}B}+h_{\textbf{k}A}d_{-\textbf{k}B})g(-\textbf{k})\right ] +h.c.
\end{align}
\end{widetext}
and the constant energy term $H_{const} = 6Nt(\chi_b\chi_f + \Delta_b\Delta_f) + 2N(\lambda+x\mu)-12N(\chi_b^{\prime}\chi_f^{\prime}+\Delta_b^{\prime}\Delta_f^{\prime})$
where $N$ is the number of unit cells. In Eq.(\ref{Hf}) and (\ref{Hb}) we have used $\alpha=A, B$ to denote the two sublattices of the honeycomb lattice. We have defined $g(\textbf{k})\equiv 1+\exp(-i k_2)+\exp(i k_1-i k_2)$, $g_1(\textbf{k}) \equiv 2[\sin k_2-\sin k_1+ \sin (k_1-k_2)]$ and $g_2({\textbf{k}}) \equiv 2\left[\cos k_1 + \cos k_2 + \cos (k_1-k_2)\right ]$ with $k_i=\textbf{k} \cdot \textbf{a}_i$.  {\em We stress that both spinon and bosonic Hamiltonians have generic hopping terms and pairing terms and they are not identical to the pairing of physical electrons, as we will explain later in our paper.}

After a mean-field approximation, it is straightforward to solve the spinon Hamiltonian \eqref{Hf} and the bosonic Hamiltonian \eqref{Hb} to obtain the ground state energy at zero temperature. Self-consistency equations are obtained via the first derivative of the ground state energy with respect to various order parameters. (See the appendix for details.) However, one has to consider possible Bose-Einstein condensations when dealing with a bosonic Hamiltonian. Since we have assumed that the translation symmetry remains unbroken, we expect a Bose-Einstein condensation can only take place at $\textbf{k}=0$. Bearing this in mind, we can explicitly separate the $\textbf{k}=0$ term from the $\textbf{k} \neq 0$ terms. We define $d_{\alpha} \equiv  1 / \sqrt{N}\langle d_{\textbf{k}=0,\alpha}  \rangle$ and $h_{\alpha} \equiv 1 / \sqrt{N}\langle h_{\textbf{k}=0,\alpha} \rangle$, both of which can acquire finite values in a condensed phase of the bosons.  Thus, $d_{\alpha}^2$ and $h_{\alpha}^2$ are the fraction of doublons and holons on sublattice $\alpha=A,B$.

We will also consider phases that break certain symmetries (lattice rotational symmetry, for example) at large $U$. This allows us to make connections to an antiferromagnetic state, which is difficult to  capture within the slave-boson mean-field approach to the Kane-Mele-Hubbard model.\cite{Vaezi:10}  In the next section we turn to a detailed description of the results of our mean-field study.  We find many features reminiscent of previous studies at half-filling, but we also obtain new results for the doped case.

\section{Mean-field results}
\label{sec:results}

In this section we discuss our mean-field results for the cases of half-filling and doping away from half-filling.  We begin with the half-filled case.

\subsection{Half-filling case}
For the case of half-filling our results are summarized in Fig. \ref{phase-diagram-at-half-filling}.  There are many similarities with results obtained in the literature via different techniques.\cite{Meng:nat10,Assaad:prl11,Wu:10,Imada:prb11,Hur:11,Rachel:prb10}  Most importantly, we find a gapped spin-liquid phase at intermediate coupling that extends to finite spin-orbit coupling.  In order of increasing interactions, the phases we find are:

\begin{figure}[tbp]
\begin{center}
\includegraphics[width=\linewidth]{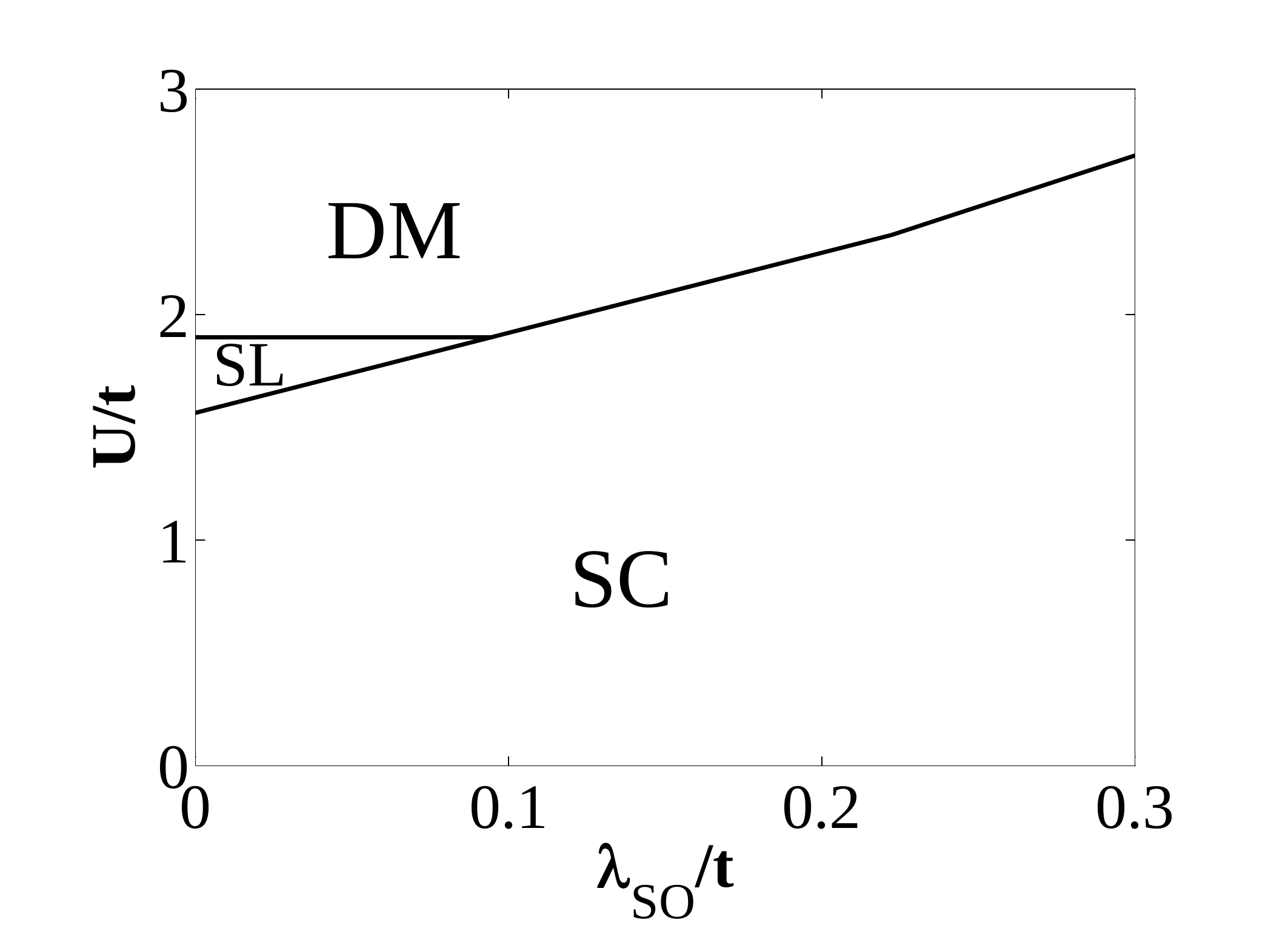}
\caption{Phase diagram of Kane-Mele-Hubbard model at half filling within the slave-boson mean-field approach. SC stands for the superconducting phase, SL is the spin liquid phase and DM is the dimerized phase. The DM phase may be considered as the closest relative to the spin density wave (SDW)/antiferromagnetic state obtained in previous studies.\cite{Meng:nat10,Assaad:prl11,Wu:10,Imada:prb11,Hur:11,Rachel:prb10}  Importantly, the slave-boson treatment also produces a gapped spin liquid at intermediate coupling (which was found in earlier numerical studies\cite{Meng:nat10,Assaad:prl11,Wu:10,Imada:prb11,Hur:11}).  We note, however, that the slave-boson treatment does not smoothly connect to the non-interacting limit since it predicts a SC phase rather than a TBI.  This is a shortcoming of the slave-boson mean-field method which is designed to capture the physics of intermediate U values where the SL phase appears.}
\label{phase-diagram-at-half-filling}
\end{center}
\end{figure}

(1) {\it Superconducting states} (SC)--When the interaction strength $U$ is small, there is a finite probability of double occupancy and empty occupancy at each site. Therefore, we expect Bose-Einstein condensation of holons and doublons could take place for small $U$. Indeed, we find a critical interaction strength $U_c(\lambda_{\rm{SO}})$ and $U_c \approx 1.5t$ at $\lambda_{\rm{SO}}=0$ above which the SC phase does not survive, which is about half of the value that has been reported in the quantum Monte Carlo simulation.\cite{Meng:nat10} (Although in that case it is a semi-metal that persists up to a critical interaction strength.) The SC phase persists even for a negative interaction, as one might expect.  One can view the $U>0$ SC region as an ``extension" from the $U<0$ region to ``small" repulsive interactions.  Recent arguments have shown that SC can, surprisingly, be expected even for (small) repulsive interactions.\cite{Raghu:prb11}  However, as we emphasized earlier, the slave-boson mean-field treatment does not properly capture the small $U>0$ physics properly in the model {\em at half-filling}.   We do expect SC states to be likely for small U upon doping, and for that reason also discuss the technical details of the half-filled case here which will only be slightly modified upon doping.

From mean-field self-consistency equations, we find that this superconducting state can be described by four finite condensates $h_A$, $h_B$, $d_A$ and $d_B$ and finite $\Delta_f$, $\Delta_b$, $\Delta_f^{\prime}$ and $\Delta_b^{\prime}$ (SC I). The four condensates are related via $h_A=-h_B$ and $d_A=-d_B$ (or other equivalent configurations). All other order parameters ({\it i.e.} the $\chi$) are zero. The physical picture for small interaction is then as follows: the spinons are paired at nearest and second nearest sites and cannot hop freely on the lattice; bosons (doublons and holons) condense independently at $\textbf{k}=0$ in momentum space. The ground state has gapless charge excitations and gapped spinon excitations. In terms of the physical electrons pairing, one can show that generally,
\begin{eqnarray}
\langle c^{\dag}_{i\uparrow} c^{\dag}_{j\downarrow} \rangle
=\langle f_{i\uparrow}^{\dag}f_{j\downarrow}^{\dag} \rangle \langle h_{i}h_{j} \rangle - \langle f_{i\downarrow}f_{j\uparrow}\rangle \langle d_{i}^{\dag}d_{j}^{\dag} \rangle \nonumber \\ -\langle f_{i\uparrow}^{\dag}f_{j\uparrow} \rangle \langle h_{i}d_{j}^{\dag} \rangle + \langle f_{i\downarrow}f^{\dag}_{j\downarrow} \rangle \langle d_{i}^{\dag}h_{j}\rangle.
\end{eqnarray}

\begin{figure}[tbp]
\begin{center}
\includegraphics[width=\linewidth]{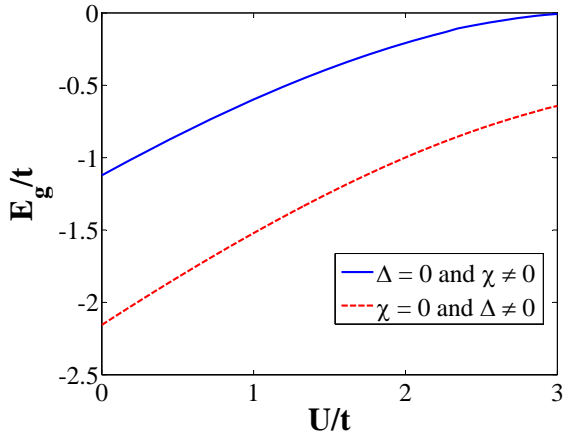}
\caption{ (Color online) The ground state energies for two slave-boson mean-field solutions.  }
\label{ground_state_energy_comparison}
\end{center}
\end{figure}

To further discuss the properties of this SC phase, let's consider singlet pairing between the same sublattices in the absence of spin-orbit coupling and we find
\begin{eqnarray}
\label{NNN-pairing-a}
\langle c^{\dag}_{0\alpha\uparrow} c^{\dag}_{\textbf{r}\alpha\downarrow} \rangle
= h_{\alpha} d_{\alpha} \sum_\textbf{k} \frac{-\lambda e^{-i \textbf{k} \cdot \textbf{r}}}{\sqrt{\lambda^2+t^2|g|^2 \Delta_b^2}},
\end{eqnarray}
where we have used fact that the Bose-Einstein condensation takes places at weak interactions so that we can replace bosonic operators with their averages. Therefore, one has finite on-site and NNN singlet pairings between same sublattices and also for neighbors arbitrarily far away. On the other hand, we find singlet pairings between different sublattices vanish. It is also possible to obtain another SC solution with finite $\chi$s and condensates but zero $\Delta$s (SC II).\cite{Vaezi:10} The spinon sector is the effective noninteracting Kane-Mele model with physical electron operators replaced by neutral spinon operators. Clearly this spinon Hamiltonian possesses non-trivial $Z_2$ topology and has time-reversal symmetry protected gapless edge states. However, singlet pairings for electrons between same sublattices are finite,
 %\begin{widetext}
%\begin{eqnarray}
%\langle c^{\dag}_{i\alpha\uparrow} c^{\dag}_{j\alpha\downarrow} \rangle= \langle h_{\alpha} \rangle \langle d_{\alpha} \rangle \left[\delta_{ij}-\sum_{\textbf{k}s=\pm1}[\Theta(-\lambda+st\chi_b|g|)]e^{ i\textbf{k} \cdot (R_j-R_i)} \right ]
%\label{onsite-pairing-b},
%\end{eqnarray}
%\end{widetext}

\begin{eqnarray}
\langle c^{\dag}_{0\alpha\uparrow} c^{\dag}_{\textbf{r}\alpha\downarrow} \rangle=  h_{\alpha}d_{\alpha} \left[\delta_{0\textbf{r}}-\sum_{\textbf{k},s=\pm1}\Theta[E_s(\textbf{k})]e^{ i\textbf{k} \cdot \textbf{r}} \right ]
\label{onsite-pairing-b},
\end{eqnarray}
where $E_s(\textbf{k}) \equiv -\lambda+st\chi_b|g(\textbf{k})|$ and $\Theta$ is the Heaviside step function. Similar to SC I, SC II has zero singlet pairings between different sublattices. That's the reason we identify it as a SC state. However, we find SC II is not energetically favorable. In Fig.~\ref{ground_state_energy_comparison}, we explicitly show the difference of two mean-field solutions. Note: our mean-field solutions at half filling only admit the above two solutions and there exists no phase with $\chi \neq 0$ and $\Delta \neq 0$ at half filling.

As pointed out in Ref.~[\onlinecite{Vaezi:10}], in the weak interacting limit the Bose gas of doublons and holons is dense (large amplitude of condensates) and one should expect the existence of strong interactions between them. Therefore, the slave-boson mean-field approach is not reliable for weak interactions. Indeed, the ground state for weak interactions in the absence (presence) of spin-orbit coupling is a Fermi liquid (TBI). This is confirmed in a recent quantum Monte Carlo study.\cite{Assaad:prl11}

Another popular approach to handle interactions, the slave-rotor mean-field approach, is believed to be able to reasonably capture the qualitative features of physics at small interactions.\cite{Georges:prb02,Georges:prb04} It has been applied to the Hubbard model on the honeycomb lattice and predicts a nodal spin liquid phase for $1.68t<U<1.74t$.\cite{Lee:prl05} Later, it was applied to Kane-Mele-Hubbard model on the same lattice and successfully predicted a TBI phase for weak interactions, though the gauge field has to be screened out to stabilize it.\cite{Rachel:prb10} The mathematical structure of slave-rotor approach allows a direct transfer of topology from physical electron bands to neutral spinons; this is the key to predicting a TBI at weak interactions and a TMI at intermediate to strong interactions. However, the slave-rotor method suffers from severe limitations for finite doping.\cite{Georges:prb02} The slave-boson mean-field approach, on the other hand, allows in principle nontrivial band topology embedded in its spinon sector. Unfortunately, in our case we obtain only finite paring terms. As a result, the slave-boson mean-field approximation falsely predicts a SC for half-filling and weak interactions. We also want to mention the Kotliar-Ruckenstein slave-boson mean-field approach describes the weak interacting limit well,\cite{Kotliar:prl86,Lilly:prl90} though it might be difficult to address the intermediate coupling regime and obtain a gapped spin liquid. It would be interesting to study its predictions and this will be left as a future work.

(2) {\it Spin liquid states} (SL)--As the interaction grows, the Bose gas becomes less dense, and one expects that the slave-boson approach is better able to describe the intermediate interaction regime. We find a spin liquid phase appears between $1.5t<U<1.9t$ for $\lambda_{\rm{SO}}=0$. In the absence of spin-orbit coupling, this phase is characterized by finite $\Delta_f$ and $\Delta_b$. Both the spinon sector and the chargeon sector are gapped and no Bose-Einstein condensation takes place. We find the singlet pairings between any two sites vanish. The expectation value of the spin at each site is also zero, and the spin-spin correlation decays exponentially due to a finite spinon gap. Therefore, we obtain a spin liquid phase in a small interaction window. Furthermore, we find it can survive over a small range of spin-orbit coupling. This feature is quite similar (even numerically) to the quantum Monte Carlo result, though the specific phase boundary differs.\cite{Assaad:prl11}

\begin{figure}[tbp]
\begin{center}
\includegraphics[width=\linewidth]{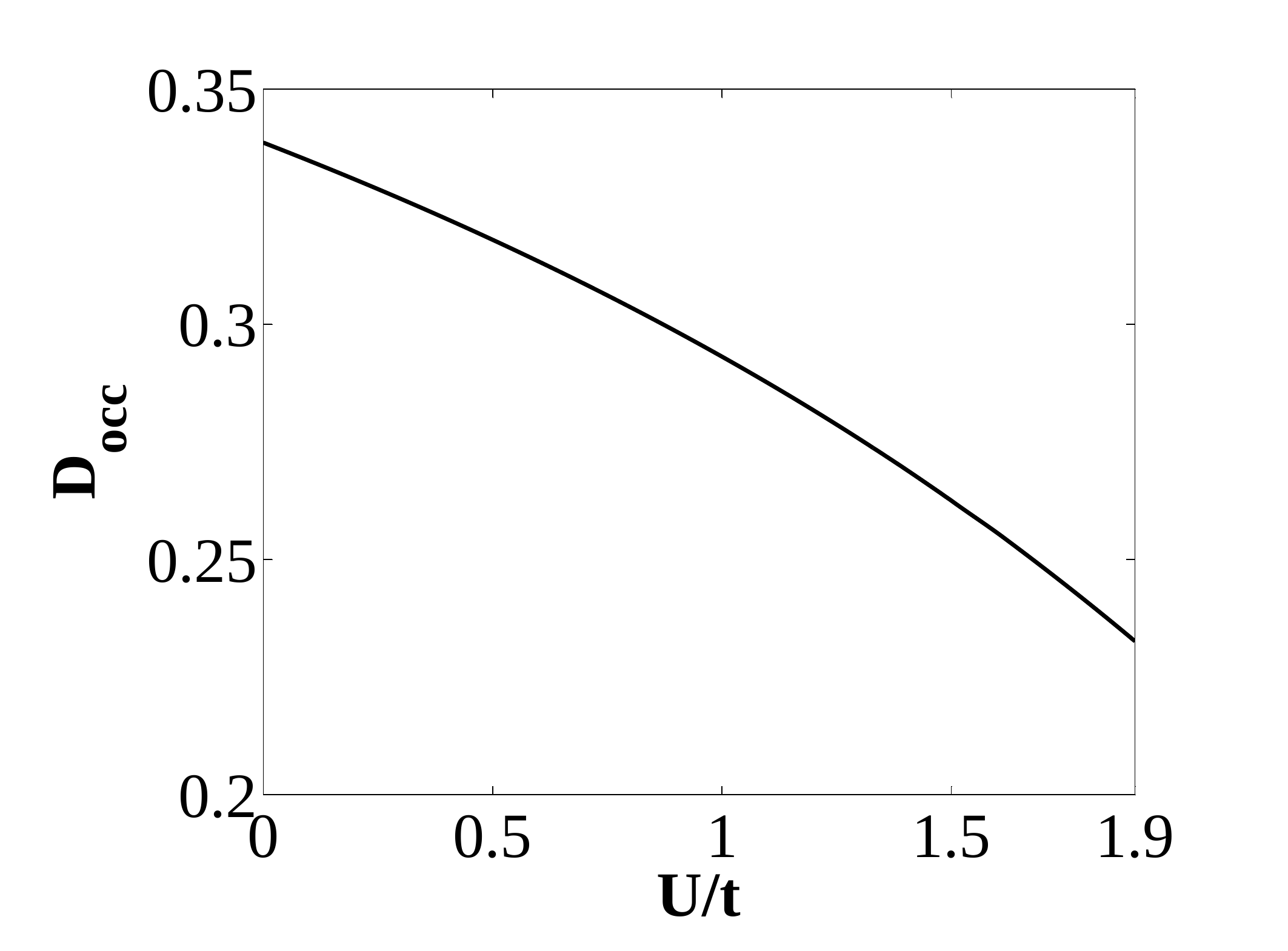}
\caption{ Double occupancy as a function of U for $\lambda_{\rm}=0.02t$ and half-filling at zero temperature from slave-boson mean-field approach. Note the double occupancy at the weak interacting limit is not correct since it is larger than $1/4$, but it gets better as the interaction grows.}
\label{double_occupancy}
\end{center}
\end{figure}

To substantiate our assertion that the slave-boson mean-field approach gets better when the interaction grows, we follow Ref.~[\onlinecite{Hur:11}] and plot the double occupancy $D_{occ} \equiv \langle n_{i\uparrow}n_{i\downarrow}\rangle$ for $\lambda_{\rm{SO}}=0.02t$ and half-filling at zero temperature in Fig.~{\ref{double_occupancy}}. As one can see, in the weak interacting regime, $D_{occ}$ is larger than 1/4 (which is the exact values for $U=0$) and this is another evidence that slave-boson mean-field approach does not work well in the weak interacting regime. However, as the interaction grows, for example, at $U=1.9t$, our $D_{occ}=0.23$ at zero temperature and this can be compared with Ref. ~[\onlinecite{Hur:11}]'s $D_{occ} \approx 0.21$ at $T=0.025t$. Since a finite temperature tends to reduce the double occupancy, we expect that our result will be very close to that of Ref.~[\onlinecite{Hur:11}]  if a zero-temperature cellular dynamical mean field study is performed.

\begin{figure}[tbp]
\begin{center}
\includegraphics[width=\linewidth]{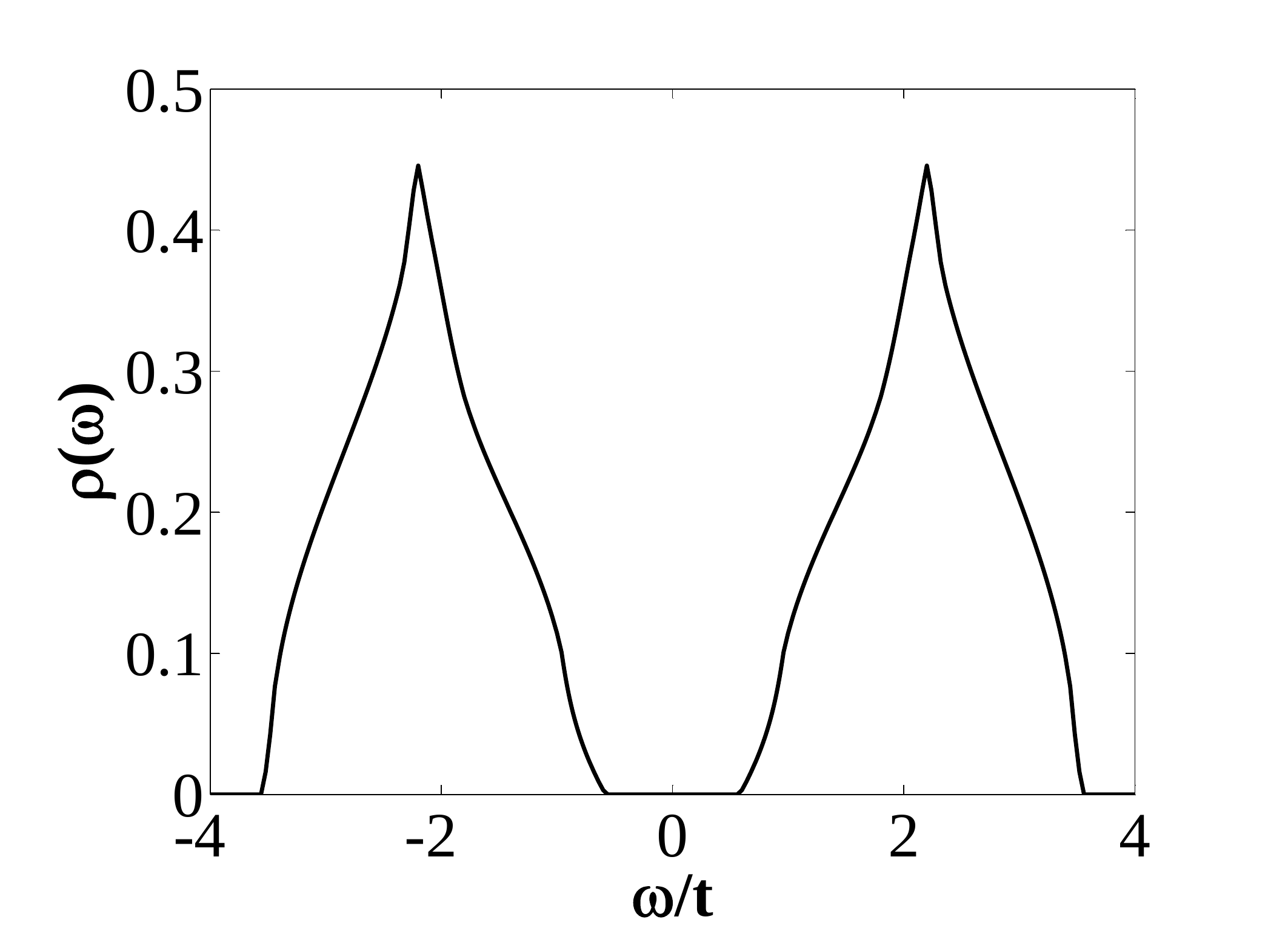}
\caption{The local density of states for the spin liquid phase at half-filling. We have taken $U=1.8t$ and $\lambda_{\rm{SO}}=0$. }
\label{local_density_of_states}
\end{center}
\end{figure}

To study the SL phase in more detail, we calculate the single particle retarded Green's function $G^{r}_{\alpha\sigma}(\textbf{k},\tau) \equiv -i\theta(t)\langle \{ c_{\textbf{k}\alpha\sigma}(t),c^{\dag}_{\textbf{k}\alpha\sigma}\} \rangle $ in the absence of spin-orbit coupling for the SL phase and the result is
%\begin{widetext}
%\begin{eqnarray}
%G^{r}_{\alpha\sigma}(\textbf{k},\omega)=
%\frac{1}{N}\sum_{\textbf{q} \neq \textbf{k}}(|v(\textbf{q-k})|^2+|V(\textbf{q})|^2) \left [ \frac{1}{\omega+i\eta-E^f(\textbf{q})-E^b(\textbf{q-k})}+\frac{1}{\omega+i\eta+E^f(\textbf{q})+E^b(\textbf{q-k})} \right ],
%\end{eqnarray}
%\end{widetext}
\begin{eqnarray}
G^{r}_{\alpha\sigma}(\textbf{k},\omega)=\sum_{\textbf{q},s=\pm1}   \frac{|v_{\textbf{q-k}}|^2+|V_{\textbf{q}}|^2}{\omega+i\eta+sE(\textbf{q},\textbf{k})},
\end{eqnarray}
where $E(\textbf{q},\textbf{k}) \equiv E^f(\textbf{q})+E^b(\textbf{q-k})$ contains a ferminonic excitation $E^f(\textbf{k}) \equiv \sqrt{\lambda^2+t^2|g(\textbf{k})|^2\Delta_b^2}$ and a bosonic excitation
$E^b(\textbf{k})\equiv \sqrt{(U/2-\lambda)^2-t^2|g(\textbf{k})|^2\Delta_f^2}$. $v_{\textbf{k}}$ and $V_{\textbf{k}}$ are defined as $|v_{\textbf{k}}|^2=\frac{1}{2}(-1+\frac{U/2-\lambda}{E^b(\textbf{k})})$ and $|V_{\textbf{k}}|^2=\frac{1}{2}(1+\frac{\lambda}{E^f(\textbf{k})})$. As we are considering half-filling, the retarded Green's function exhibits particle-hole symmetry. To make our calculations more solid, we first check if the sum rule of spectrum function $\rho(\textbf{k},\omega) \equiv \frac{1}{\pi}\textrm{Im}[G^{r}_{\alpha\sigma}]$ is satisfied. Since it is based on the anticommunication relations between $c_{\textbf{k}}$ and $c^{\dag}_{\textbf{k}} $ and it has been taken into account by Eq.~\ref{constrain-completeness} on the average, the sum rule of our slave-boson mean-field approach is implicitly fulfilled by the mean-field equations. The local density of states $\rho(\omega) \equiv \sum_{\textbf{k}} \rho(\textbf{k},\omega)$ is shown in Fig.~\ref{local_density_of_states} for $U=1.8t$ and $\lambda_{\rm{SO}}=0$. The single particle gap is found to be $0.57t$. This is the gap at the Dirac point. Instead of calculating it numerically in Ref.~\onlinecite{Meng:nat10}, we can determine it analytically in our case. The poles of the retarded Green's function are at $\omega=\pm E(\textbf{q},\textbf{k})$ and the positive pole reaches its minimum at Dirac point, $\textbf{q}=\textbf{k}=\textbf{K}$, therefore it is clear that the single particle gap at the Dirac point is $\Delta_{\textrm{sp}}=E(\textbf{K},\textbf{K})=|\lambda|+\sqrt{(U/2-\lambda)^2-9t^2\Delta_f^2}=0.57t$ for $U=1.8t$. Since our phase boundary for SL differs from Ref.~[\onlinecite{Meng:nat10}], we cannot compare the single particle gap directly for the same $U$. However, our result for a point sitting about in the middle of SL phase ($0.57t$) is comparable to a typical single particle gap from Ref.~[\onlinecite{Meng:nat10}] (about $0.1t$).

A SL is also found in the slave-rotor mean-field approach, though its properties are quite different from the one obtained here.\cite{Rachel:prb10,Kou:prb11} For example, only hopping terms of spinons are present in the SL within the slave-rotor approach and it has a U(1) gauge symmetry. In 2D, U(1) gauge fluctuations are important\cite{Witczak-Krempa:prb10} and it has been argued that other gapless layers (spinons) are required to screen the gauge field and suppress the gauge fluctuations.\cite{Young:prb08} Our spin/charge gapped SL, however, does not require an additional layer to stabilize the mean-field result. The key difference with the slave-boson approach is that the effective fermionic Hamiltonian for the SL consists of pairing terms instead of hopping of spinons. As a result, the presence of NNN pairings allow the staggered U(1) gauge symmetry to break down to a $Z_2$ gauge symmetry by the Anderson-Higgs mechanism and gap out the U(1) fluctuations.\cite{Vaezi:10} Therefore, our mean-field result has at least a chance of being realistic, and  quantum Monte Carlo calculations\cite{Assaad:prl11} and dynamical mean-field theory calculations\cite{Hur:11} support this result in a similar parameter regime.

(3) {\em Dimerized phase} (DM)--The spin liquid phase is unstable to dimerization in the large $U$ limit. The dimerized phase is close in many respects to an antiferromagnetic phase, which is expected to occur at large interactions on a bi-partite lattice like the honeycomb lattice.\cite{Meng:nat10,Assaad:prl11,Wu:10,Imada:prb11,Hur:11,Rachel:prb10}  With the present form of Kane-Mele-Hubbard model (in the absence of a spin-exchange term), it is difficult to include antiferromagnetic order in our mean-field approach.\cite{Vaezi:10} We will instead turn to a dimerized phase which has anisotropy in some direction ({\it i.e.} breaks lattice rotational symmetry) yet keeps the translational symmetry intact. Similar ideas have been applied in the slave-rotor approach.\cite{Young:prb08} To perform our calculations, we will assume rotational symmetry is spontaneously broken. We consider an ansatz of three different mean-field $\Delta_{f1}$, $\Delta_{f2}$, and $\Delta_{f3}$ for NN pairings, and $\Delta_{f1}^{\prime}$, $\Delta_{f2}^{\prime}$, and $\Delta_{f3}^{\prime}$ for NNN pairings. We do the same in the chargeon sector. In the parameter space we consider, the mean-field solutions are those that satisfy $\Delta_{f1} \neq 0$, $\Delta_{f2}=\Delta_{f3}=0$ and $\Delta_{b1} \neq 0$, $\Delta_{b2}=\Delta_{b3}=0$ while the NNN pairings for spinons and bosons vanish. This is an extreme example of dimerization and it corresponds to an atomic-like insulator which consists of noninteracting pairs of NN sites.  This can be taken as a ``closest cousin" to the antiferromagnetic state expected at large $U$ for half-filling.

\begin{figure}[tbp]
\begin{center}
\includegraphics[width=\linewidth]{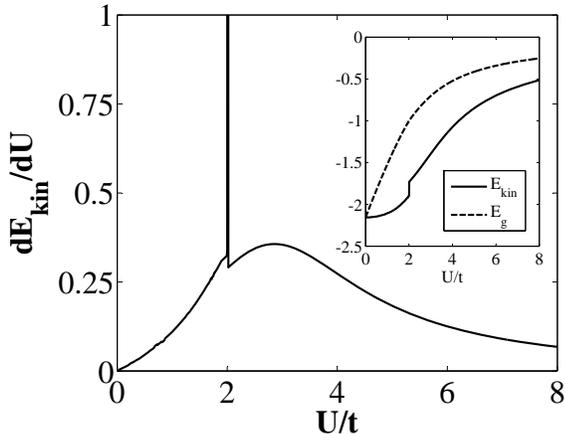}
\caption{ The derivative of kinetic energy per unit cell $dE_{kin}/dU$ for $\lambda_{\rm{SO}}=0$. Insert: the kinetic energy and the ground state energy. The kinetic energy has a jump around $U=1.9t$ and the derivative of it shows a sharp peak at the same location, which indicates a first order transition around $U=1.9t$. }
\label{Ek_and_Eg}
\end{center}
\end{figure}

To make further connections to the numerical studies, we follow Ref~[\onlinecite{Meng:nat10}] and plot the derivative of the kinetic energy per unit cell $dE_{kin}/dU$ as a function of $U/t$ in Fig.~\ref{Ek_and_Eg}. After a comparison to the QMC, we find: (i) our kinetic energy is higher than the one in QMC and we expect our ground state energy is also higher, though we are not aware of reported ground state energy in QMC; (ii) Our kinetic energy profile resembles the one in QMC, though we have a jump around $U=1.9t$ from the spin liquid phase to the dimer phase while it has a continuous behavior in QMC; (iii) Our calculations show that we have a second order phase transition at the first critical point $U_{c1}=1.5t$ followed by a first order phase transition at $U_{c2}=1.9t$ while there appears to be a continuous Mott transition around $U=3.5t$.\cite{Wang:prb10}

\subsection{Doping Cases}
Up to this point, we have focused on the case of half-filling and our mean-field results could be directly compared with other numerical approaches.\cite{Meng:nat10,Assaad:prl11,Wu:10,Imada:prb11,Hur:11,Rachel:prb10}  We now break new ground by considering the case of doping away from half-filling where other methods may encounter serious shortcomings.

The doped Hubbard model in the strongly interacting limit and its descendant $t$-$J$ model are believed to capture the physics of high temperature superconductivity upon doping.\cite{Wen:rmp06,Senthil:prb05} In most slave-boson treatments, one assumes strong interactions and simplifies the calculations by removing double occupancy from the Hilbert space. However, since we are mostly interested in the intermediate regime where $U$ and $t$ are comparable, and a possible spin liquid phase resides, we will start with the Kane-Mele-Hubbard model without assuming a strong interaction. We therefore retain the entire Hilbert space.

Using the mean-field self-consistency equations (see Appendix for details), one finds that the SL at half-filling is unstable to infinitesimal doping and a Bose-Einstein condensations of chargeons takes place for {\em any} doping. This can be seen from Eq.~(\ref{eq-x}) where the doping is directly related to the condensates. The number of doublons at each site is not equal to the number of holons, and at least one of them has to be finite. This indicates the onset of Bose-Einstein condensation for any doping. Our mean-field solutions show that the $\chi$s also acquire finite values, \emph{i.e.} spinons and chargeons can both hop and form pairs on the lattice.

\begin{figure}[tbd]
\begin{center}
\includegraphics[width=\linewidth]{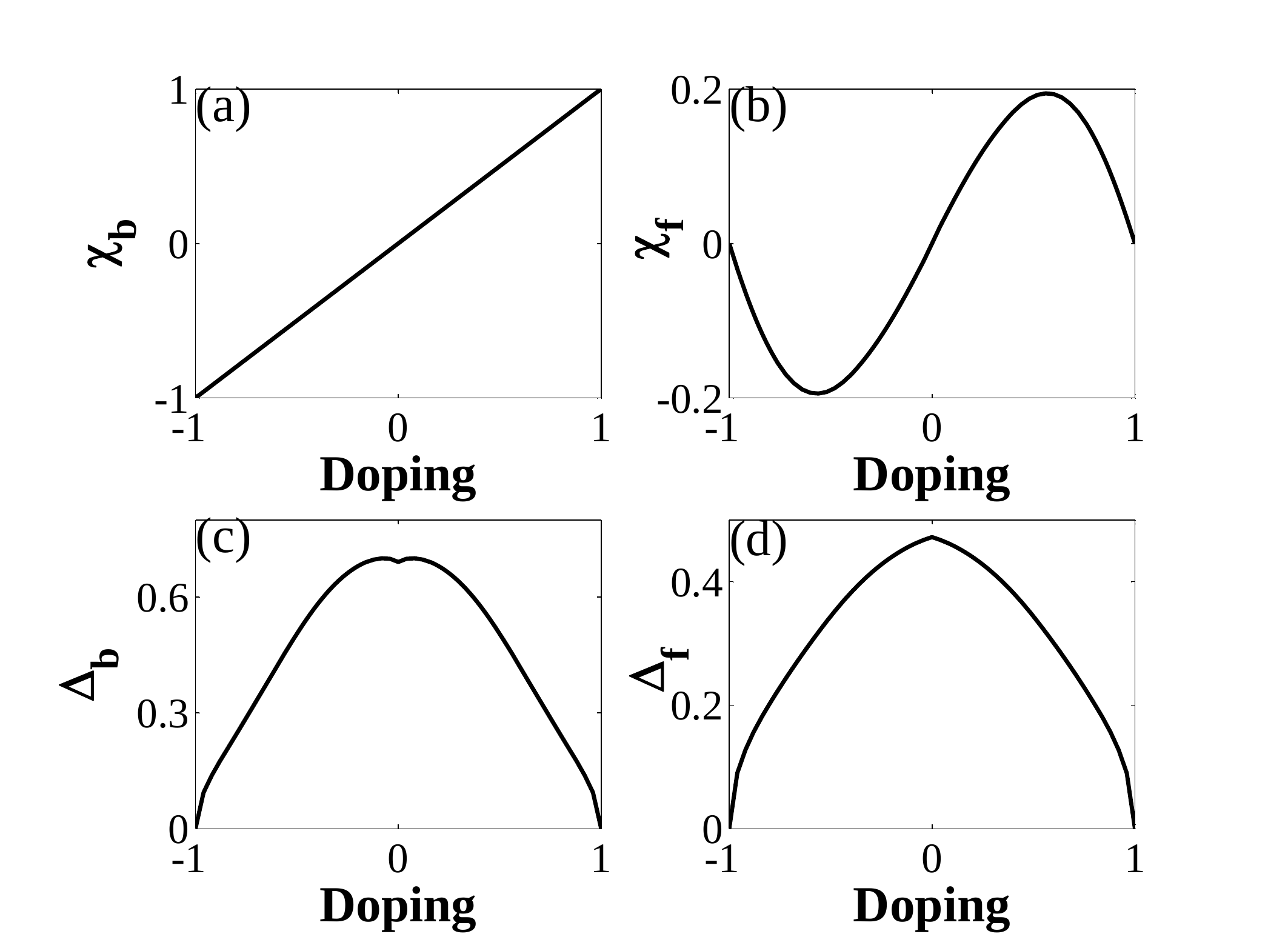}
\caption{The NN order parameters as functions of doping. We have set $U=1.8t$ and $\lambda_{\rm{SO}}=0.05t$, which is a SL at half-filling.}
\label{NN_order_parameters}
\end{center}
\end{figure}

\begin{figure}[tbd]
\begin{center}
\includegraphics[width=\linewidth]{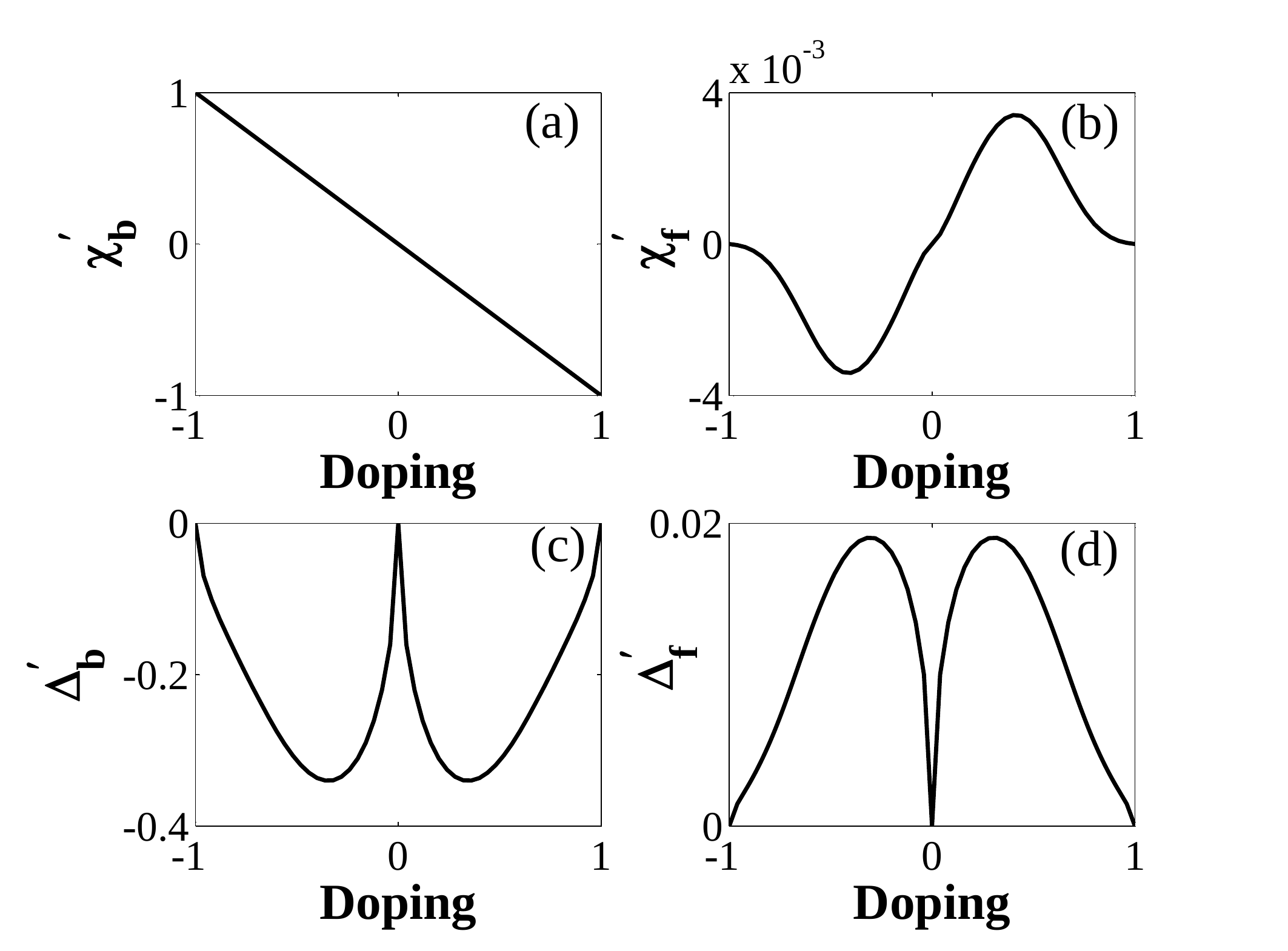}
\caption{The NNN order parameters as functions of doping. We have set $U=1.8t$ and $\lambda_{\rm{SO}}=0.05t$, which is a SL at half-filling.}
\label{NNN_order_parameters}
\end{center}
\end{figure}

In Fig.~\ref{NN_order_parameters} and Fig.~\ref{NNN_order_parameters}, we show various NN and NNN order parameters. As one can see, $\chi_b$ and $\chi_b^{\prime}$ have linear relations with respect to doping, which readily follows from Eq.~(\ref{eq-x}), Eq.~(\ref{eq-chib}) and Eq.~(\ref{eq-chibp}). $\chi_f$ and $\chi_f^{\prime}$ have similar behaviors and are odd functions of doping while $\Delta_b$, $\Delta_f$,$\Delta_b^{\prime}$ and $\Delta_f^{\prime}$ are even functions of doping. Interestingly, the value of $\Delta_b^{\prime}$ and $\Delta_f^{\prime}$ are numerically very close to zero at half filling. The four condensates are related via $h_A=-h_B$ and $d_A=-d_B$(or other equivalent configurations).

\begin{figure}[tbp]
\begin{center}
\includegraphics[width=\linewidth]{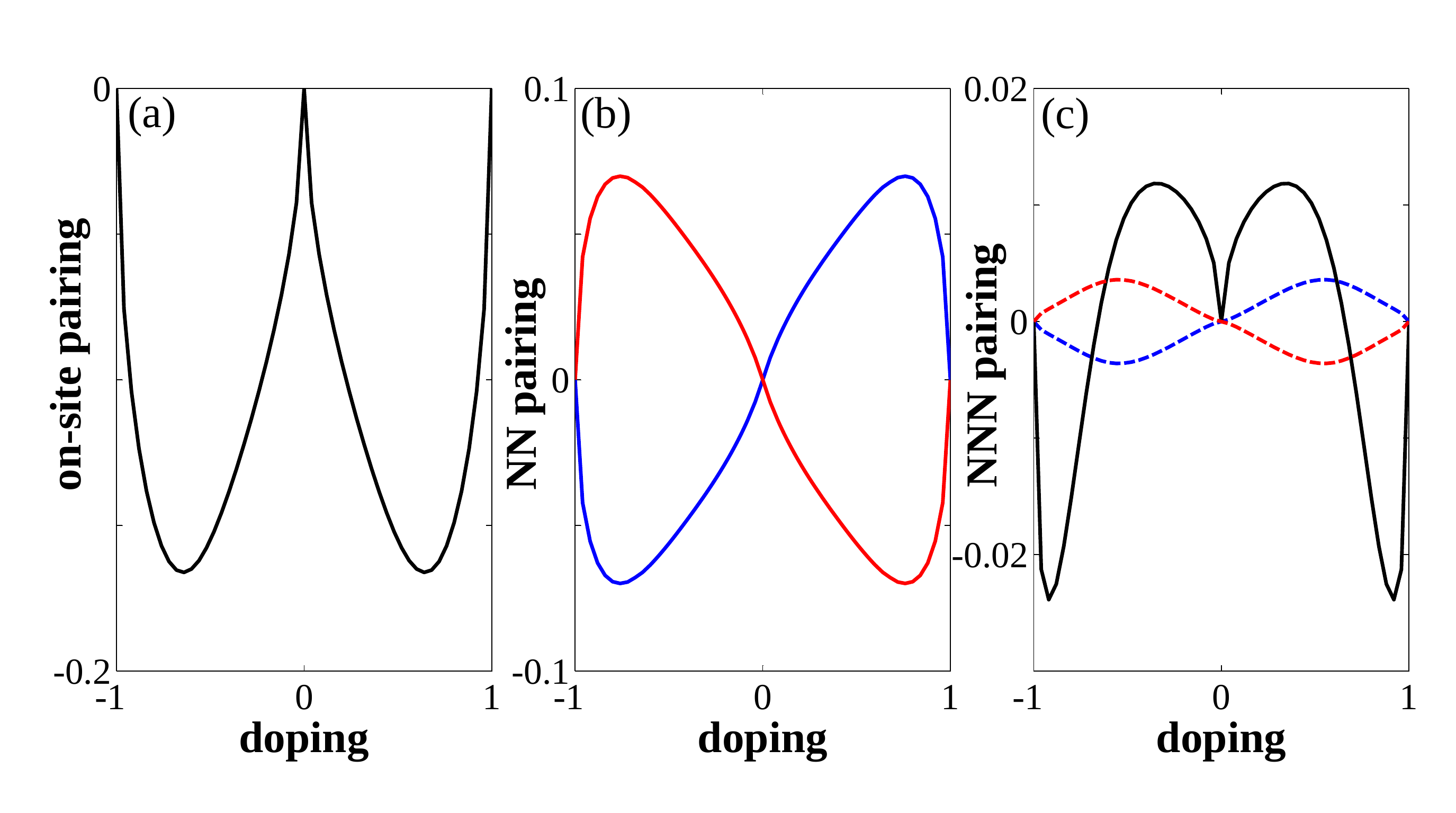}
\caption{(Color online) Singlet parings at $U=1.8t$ and $\lambda_{\rm{SO}}=0.05t$ corresponding to the SL at half-filling in Fig.\ref{phase-diagram-at-half-filling}. Shown as a function of doping (additional electrons/site) is: (a) The on-site pairing, (b) the nearest neighbor paring, and (c) the next nearest neighbor pairing. The black solid line is for the real part of the pairing, which is identical for both sublattices; the blue and red solid (dash) lines are for the real (imaginary) part of paring that is different for A and B sublattices.}
\label{singlet_pairing_at_doping}
\end{center}
\end{figure}

In Fig. \ref{singlet_pairing_at_doping}, we plot the physical onsite, NN and NNN singlet pairings as a function of doping for parameters $U=1.8t$ and $\lambda_{\rm{SO}}=0.05t$, whose ground state is a SL without doping. As the doping is increased, singlet pairings between same sublattices and different sublattices acquire finite values and signal the occurrence of a SC phase. The singlet pairings are not monotonic functions of the doping and there exists an ``optimal" doping (around $\pm 0.8$ electron/site) where the magnitude of on-site and NN parings are maximized. This bears some similarity to the famous SC ``dome" in the phase diagram of high temperature superconductors,\cite{Wen:rmp06} though electron doping and hole doping are ``equivalent" in our case. We also remark that the dimerized phase will become a SC state via doping. Therefore, upon doping the SC phase takes over the entire phase diagram within the slave-boson mean-field treatment.  However, as we mentioned earlier, the SC phase obtained is not one that possess topological order of any obvious type.

\begin{figure}[tbd]
\begin{center}
\includegraphics[width=\linewidth]{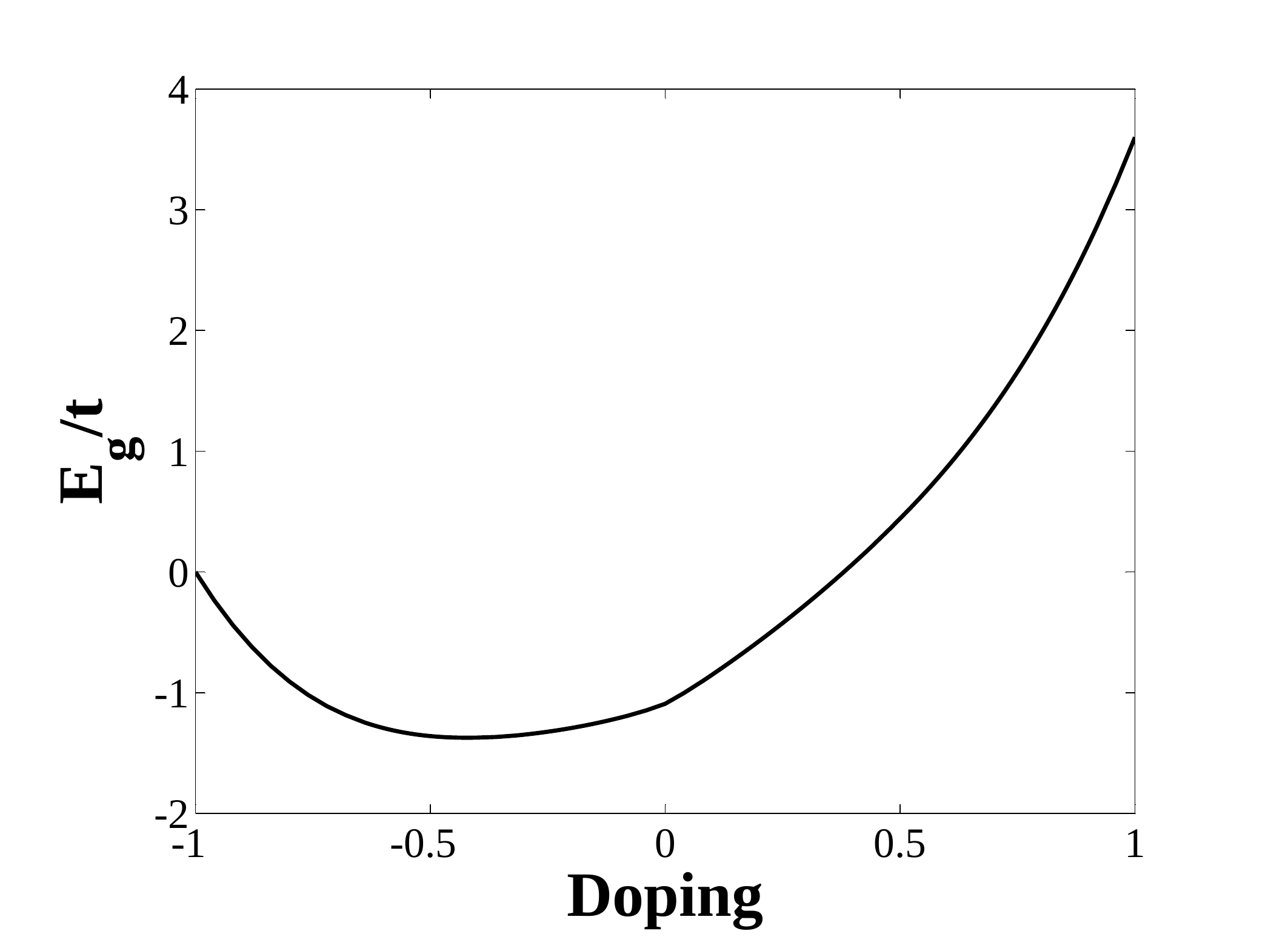}
\caption{The ground state energy per unit cell $E_g$ as a function of doping. We have set $U=1.8t$ and $\lambda_{\rm{SO}}=0.05t$, which is a SL at half-filling.}
\label{Eg_at_doping}
\end{center}
\end{figure}

In Fig.~\ref{Eg_at_doping}, we plot the ground state energy as a function of doping. We have set $U=1.8t$ and $\lambda_{\rm{SO}}=0.05t$, which is a SL at half-filling. For the ground state energy, one can understand it as follows. At $x=-1$ where electrons are completely depleted the energy is zero, and when one starts to add more electrons
to the system, the ground state energy decreases since kinetic energy dominates over the potential energy and lowers the ground state energy. As more electrons are added, the potential energy starts to dominate and cause the increase of ground state energy. This happens around $x=-0.5$ where $\chi_f$ (a measure of kinetic energy) acquires the maximum amplitude. Eventually, when the number of electrons reaches 2 per site, electrons are frozen at each sites and they cannot hop any more and the ground state energy is the classical potential energy ($3.6t$ in our case). Therefore, our slave-boson mean-field calculations is able to replicate the exact ground state energy at two doping limits, and we expect it should describe the intermediate doping well. We comment our ground state energy bears a similar trend to the one in Kotliar-Ruckenstein slave-boson mean-field approach. \cite{Lilly:prl90}

\section{Conclusions}
\label{sec:conclusions}

In this paper we have studied the Kane-Mele-Hubbard model on the honeycomb lattice via the slave-boson mean-field approach. We have considered both the case of half-filling, which has been addressed earlier in the literature via other methods, and the case of doping, which has not been previously investigated to the best of our knowledge.  Our main results are summarized in Fig. \ref{phase-diagram-at-half-filling} and Fig. \ref{singlet_pairing_at_doping}.

At half-filling, the slave-boson mean-field approach fails to capture the correct physics of weak interactions and predicts a SC state (rather than a TBI), but we find it correctly predicts a spin liquid phase for intermediate interactions and small spin-orbit coupling.  This is one of the least well understood regimes, and in the presence of strong spin-orbit coupling is likely to lead to new phases.\cite{Pesin:np10,Kargarian:prb11,Ruegg11}  It would be interesting to consider models with further range interactions (first or second-neighbor) to see if they might favor any new phases in the phase diagram, and possibly other mean-field ansatz for the present case as well.

With finite doping, the spin liquid and dimerized phases become superconducting states  characterized by finite singlet parings (and the superconducting state at half-filling remains a superconducting state).  Unfortunately,  all the superconducting states we find do not possess any obvious form of topological order.  Thus, our work leaves largely open the question of how likely doping-induced topological superconducting states are to appear in models that support interacting topological insulators at half-filling.  We hope our work will help to stimulate future studies on the effects of doping topological insulators, including those with longer-range interactions.  Doping three dimensional multi-orbital models also seems a promising direction.\cite{Pesin:np10,Kargarian:prb11}

\section{acknowledgement}
We thank Andreas R\"uegg for enlightening discussions. We gratefully acknowledge financial support from ARO Grant W911NF-09-1-0527 and NSF Grant DMR-0955778.

\appendix
\section{Slave-boson self-consistency equations}
\label{app:formulas}

In this section, we provide some details on the mean-field self-consistency equations and order parameters for readers interested in the details of our calculations. To obtain the self-consistency equations, we start with the ground state energy per unit cell $E_g = E_f+E_b+E_{c}$
where $E_f$ ($E_b$) is the ground state energy from the fermionic (bosonic) part and $E_c$ is an energy constant. We have
\begin{equation}
\label{Ef}
E_f = - \frac{1}{N}\sum_{\textbf{k}}\left[\left( A_1 -2 \sqrt{A_2}\right)^{1/2} + \left(A_1 +2 \sqrt{A_2}\right)^{1/2}\right],
\end{equation}
where $A_1$ and $A_2$ are defined as
\begin{widetext}
\begin{equation}
A_1 \equiv \lambda^2+|g|^2 t^2 \Delta_b^2+g_1^2 \Delta_{b}^{\prime 2} \lambda_{\rm{SO}}^2+|g|^2 t^2 \chi_b^2+g_1^2 \lambda_{\rm{SO}}^2 \chi_{b}^{\prime 2},
\end{equation}

\begin{equation}
A_2 \equiv  |g|^2 \lambda^2 t^2 \chi _b^2+|g|^2 t^2 g_1^2 \Delta_{b}^{\prime 2} \lambda_{\rm{SO}}^2 \chi _b^2-2 |g|^2 t^2 g_1^2 \Delta _b \Delta_{b}^{\prime} \lambda_{\rm{SO}}^2 \chi _b \chi_{b}^{\prime}+\lambda^2 g_1^2 \lambda_{\rm{SO}}^2 \chi_{b}^{\prime 2}+|g|^2 t^2 g_1^2 \Delta _b^2 \lambda_{\rm{SO}}^2 \chi_{b}^{\prime 2}.
\end{equation}

The bosonic ground state energy is
\begin{eqnarray}
\label{Eb}
E_b = -U+\frac{1}{2N}\sum_{\textbf{k}}\left [\sqrt{(-2 \lambda+U)^2-4 |g t \Delta _f- g_2 \Delta_{f}^{\prime} \lambda_{\rm{SO}}|^2}+\sqrt{(-2 \lambda+U)^2-4 |g t \Delta _f+ g_2 \Delta_{f}^{\prime} \lambda_{\rm{SO}}|^2} \right ],
\end{eqnarray}
where we have chosen excitation spectra\cite{RipkaBook:1986} that may give rise to Bose-Einstein condensation at $\textbf{k}=0$. The energy
\begin{eqnarray}
\label{Ec}
E_c &=  2 \lambda+2 x \mu -6 (d_B h_A+d_A h_B) t \Delta _f+12 (d_A h_A+d_B h_B) \Delta_{f}^{\prime} \lambda_{\rm{SO}}+6 (d_A d_B-h_A h_B) t \chi _f
+6 t \left(\Delta _b \Delta _f+\chi _b \chi _f\right) \nonumber \\ &+\left(h_A^2+h_B^2\right) \left(-\lambda+\mu +6 \lambda_{\rm{SO}} \chi_{f}^{\prime}\right) -\left(d_A^2+d_B^2\right) \left(\lambda-U+\mu +6 \lambda_{\rm{SO}} \chi_{f}^{\prime}\right)-12 \lambda_{\rm{SO}}(\Delta_{b}^{\prime} \Delta_{f}^{\prime}+\chi_{b}^{\prime} \chi_{f}^{\prime}).
\end{eqnarray}

Taking the derivative of $E_g$ with respect to the order parameters, we immediately obtain the self-consistency equations:
\begin{equation}
\label{eq-x}
x=\frac{1}{2}(d_A^2+d_B^2-h_A^2-h_B^2),
\end{equation}

\begin{equation}
\label{eq-chib}
\chi_b = h_A h_B - d_A d_B,
\end{equation}

\begin{equation}
\label{eq-chibp}
\chi_{b}^{\prime}=-\frac{1}{2}(d_A^2+d_B^2-h_A^2-h_B^2),
\end{equation}

\begin{eqnarray}
 {d_A}( \lambda- U+ \mu +6\lambda_{\rm{SO}} \chi_{f}^{\prime})+3  {h_B} t \Delta _f-6  {h_A} \Delta_{f}^{\prime} \lambda_{\rm{SO}}-3  {d_B} t \chi _f=0, \\
 d_B(\lambda - U +\mu +6 \lambda_{\rm{SO}} \chi_{f}^{\prime})  +3  {h_A} t \Delta _f-6  {h_B} \Delta_{f}^{\prime} \lambda_{\rm{SO}}-3  {d_A} t \chi _f =0,\\
 {h_A}( -\lambda+  \mu +6  \lambda_{\rm{SO}} \chi_{f}^{\prime}) -3  {d_B} t \Delta _f+6  {d_A} \Delta_{f}^{\prime} \lambda_{\rm{SO}}-3  {h_B} t \chi _f =0, \\
 {h_B}( -\lambda+ \mu +6 \lambda_{\rm{SO}} \chi_{f}^{\prime})-3  {d_A} t \Delta _f+6  {d_B} \Delta_{f}^{\prime} \lambda_{\rm{SO}}-3  {h_A} t \chi _f=0,
\end{eqnarray}

\begin{eqnarray}
\label{eq-chif}
\chi_f = \frac{1}{6Nt}\sum_{\textbf{k}} \left[  \left(2|g|^2 t^2 \chi_b-\frac{d_6}{d_5}\right)/d_1 +\left(2|g|^2 t^2 \chi_b + \frac{d_6}{d_5}\right)/d_2 \right],
\end{eqnarray}
\begin{eqnarray}
\label{eq-Deltaf}
\Delta_f = \frac{1}{6Nt}\sum_{\textbf{k}} \left [    \left(2|g|^2 t^2 \Delta_b+ \lambda_{\rm{SO}}^2 \frac{d_7}{d_5}\right )/d_1 +\left(2 |g|^2 t^2 \Delta_b- \lambda_{\rm{SO}}^2\frac{d_7}{d_5}\right)/d_2  \right],
\end{eqnarray}
\begin{equation}
\label{eq-Deltab}
\Delta_b = d_A h_B + h_A d_B + \frac{1}{3N}\sum_{\textbf{k}}\bigg[  |g|  \big( \left(|g| t \Delta_f+g_2 \Delta_{f}^{\prime} \lambda_{\rm{SO}} \right)/d_3
      + \left(|g| t \Delta_f-g_2 \Delta_{f}^{\prime} \lambda_{\rm{SO}} \right) /d_4 \big )\bigg],
\end{equation}

\begin{equation}
\label{eq-Deltafp}
\Delta_{f}^{\prime}=-\frac{\lambda_{\rm{SO}}}{12N}\sum_{\textbf{k}}\left[   \left(2 g_1^2 \Delta_{b}^{\prime}-\frac{d_{10}}{d_5}\right)/d_1+\left(2 g_1^2 \Delta_{b}^{\prime}+\frac{d_{10}}{d_5} \right)/d_2   \right],
\end{equation}

\begin{equation}
\label{eq-Deltabp}
\Delta_{b}^{\prime}=h_A d_A + h_B d_B-\frac{1}{6N}\sum_{\textbf{k}}\bigg[   g_2\big(
            - \left( |g|t\Delta_f+g_2\Delta_{f}^{\prime} \lambda_{\rm{SO}}\right)/d_3
            + \left( |g|t\Delta_f-g_2\Delta_{f}^{\prime} \lambda_{\rm{SO}}\right)/d_4\big) \bigg],
\end{equation}

\begin{equation}
\label{eq-chifp}
\chi_{f}^{\prime}=-\frac{\lambda_{\rm{SO}}}{12N}\sum_{\textbf{k}}\bigg[
              \big(2 g_1^2 \chi_{b}^{\prime}-\frac{d_8} {d_5} \big)/d_1
            +\big (2 g_1^2 \chi_{b}^{\prime}+ \frac{d_8} {d_5} \big)/d_2
 \bigg ],
\end{equation}

\begin{eqnarray}
\label{eq-lambda}
2=d_A^2+d_B^2+h_A^2+h_B^2-\frac{1}{2N}\sum_{\textbf{k}}\bigg[   (4 \lambda-2U) \left(\frac{1}{d_3} +\frac{1}{d_4} \right )+ 2\left(-2\lambda+\frac{d_9}{d_5}\right)/d_1- 2\left(2\lambda+\frac{d_9}{d_5}\right)/d_2  \bigg],
\end{eqnarray}
where the $d_i$ are defined as follows:
\begin{eqnarray}
d_{1,2}=2 \bigg[ \lambda^2+|g|^2 t^2 \left(\Delta _b^2+\chi _b^2\right)+g_1^2 \lambda_{\rm{SO}}^2 \left(\Delta_{b}^{\prime 2}+\chi_{b}^{\prime 2}\right) \nonumber
\\ \mp 2 \sqrt{|g|^2 t^2 \left(\lambda^2+g_1^2 \Delta_{b}^{\prime 2} \lambda_{\rm{SO}}^2\right) \chi _b^2-2 |g|^2 t^2 g_1^2 \Delta _b \Delta_{b}^{\prime} \lambda_{\rm{SO}}^2 \chi _b \chi_{b}^{\prime}+g_1^2 \left(\lambda^2+|g|^2 t^2 \Delta _b^2\right) \lambda_{\rm{SO}}^2 \chi_{b}^{\prime 2}}\bigg]^{\frac{1}{2}},
\end{eqnarray}

\begin{eqnarray}
d_{3,4}=\sqrt{(-2 \lambda+U)^2-4 |g|^2 t^2 \Delta _f^2 \mp 8 |g| t g_2 \Delta _f \Delta_{f}^{\prime} \lambda_{\rm{SO}}-4 g_2^2 \Delta_{f}^{\prime 2} \lambda_{\rm{SO}}^2},
\end{eqnarray}

\begin{eqnarray}
d_5 =\sqrt{|g|^2 t^2 \left(\lambda^2+g_1^2 \Delta_{b}^{\prime 2} \lambda_{\rm{SO}}^2\right) \chi _b^2-2 |g|^2 t^2 g_1^2 \Delta _b \Delta_{b}^{\prime} \lambda_{\rm{SO}}^2 \chi _b \chi_{b}^{\prime}+g_1^2 \left(\lambda^2+|g|^2 t^2 \Delta _b^2\right) \lambda_{\rm{SO}}^2 \chi_{b}^{\prime 2}},
\end{eqnarray}

\begin{eqnarray}
d_6=2 |g|^2 t^2 \left[\left(\lambda^2+g_1^2 \Delta_{b}^{\prime 2} \lambda_{\rm{SO}}^2\right) \chi _b-g_1^2 \Delta _b \Delta_{b}^{\prime} \lambda_{\rm{SO}}^2 \chi_{b}^{\prime}\right],
\end{eqnarray}

\begin{eqnarray}
d_7=2 |g|^2 t^2 g_1^2  \chi_{b}^{\prime} \left(\Delta_{b}^{\prime} \chi _b-\Delta _b \chi_{b}^{\prime}\right),
\end{eqnarray}

\begin{eqnarray}
d_8=2 g_1^2  \left(-|g|^2 t^2 \Delta _b \Delta_{b}^{\prime} \chi _b+\lambda^2 \chi_{b}^{\prime}+|g|^2 t^2 \Delta _b^2 \chi_{b}^{\prime}\right),
\end{eqnarray}

\begin{eqnarray}
d_9=2 \lambda \left(|g|^2 t^2 \chi _b^2+g_1^2 \lambda_{\rm{SO}}^2 \chi_{b}^{\prime 2}\right),
\end{eqnarray}
and
\begin{eqnarray}
d_{10}=2 |g|^2 t^2 g_1^2 \chi _b \left(\Delta_{b}^{\prime} \chi _b-\Delta _b \chi_{b}^{\prime}\right).
\end{eqnarray}
\end{widetext}

%\bibliography{slave-boson_paper-6-30-2011}
\bibliography{biblio,top_ins,top_ins_rev}
\end{document}